\documentclass{acm_proc_article-sp}

\usepackage{booktabs}

\begin{document}

\title{Document Distance for the Automated Expansion of Relevance
  Judgements for Information Retrieval Evaluation}


%
%
%
%

\numberofauthors{3} 
%
\author{
%
%
\alignauthor
Diego Moll\'a\\
       \affaddr{Department of Computing}\\
       \affaddr{Macquarie University}\\
       \affaddr{Sydney, Australia}\\
       \email{\normalsize diego.molla-aliod@mq.edu.au}
\alignauthor
Iman Amini\\
       \affaddr{NICTA and}\\
       \affaddr{RMIT}\\
       \affaddr{Melbourne, Australia}\\
       \email{\normalsize iman.amini@rmit.edu.au}
\alignauthor
David Martinez\\
       \affaddr{University of Melbourne}\\
       \affaddr{Melbourne, Australia}\\
       \email{\normalsize davidm@csse.unimelb.edu.au}
}



\toappear{\the\boilerplate\par
{\confname{\the\conf}} \the\confinfo\par \the\copyrightetc}

\permission{Copyright is held by the author/owner(s).}
\conferenceinfo{SIGIR'14 Workshop on Gathering Efficient Assessments of Relevance (GEAR'14),}{\\July 11, 2014, Gold Coast, Queensland, Australia.}
\copyrightetc{}


\maketitle

\begin{abstract} 
  This paper reports the use of a document distance-based approach to
  automatically expand the number of available relevance judgements
  when these are limited and reduced to only positive judgements. This
  may happen, for example, when the only available judgements are
  extracted from a list of references in a published review paper. We
  compare the results on two document sets: OHSUMED, based on medical
  research publications, and TREC-8, based on news feeds. We show that
  evaluations based on these expanded relevance judgements are more
  reliable than those using only the initially available judgements,
  especially when the number of available judgements is very limited.
\end{abstract}




\section{Introduction}

An important bottleneck in the development of information retrieval
(IR) systems is their evaluation. Generating human-produced judgements
is expensive and time-consum\-ing, and it is not always possible to
produce a large set of relevance judgements (qrels henceforth).

We envisage a scenario where the only available qrels are the list of
references of a survey paper. For example, within the area of Evidence
Based Medicine (EBM), clinical systematic reviews provide the key
published evidence that is relevant to a specific clinical query,
together with a list of references that backs up the clinical
evidence. This list of references, however, covers only a small sample
of all relevant references~\cite{Dickersin:1994}. Furthermore, only a
fraction of the documents of a systematic review can be retrieved
after performing exhaustive searches, mostly due to the fact that
there are complex queries and several document
repositories~\cite{martinez2008}. Another problem with using the list
of references as the only qrels is that negative qrels, that is,
judgements about non-relevant documents, are not included. Any
attempts to develop IR systems for such a scenario will need to
supplement the list of references with something else. In this paper
we propose to automatically expand the qrels by finding similar
documents.

\section{Related Work}

Using document distance as a criterion to expand a list of qrels
sounds intuitive. The approach is related to the well-known cluster
hypothesis: ``closely associated documents tend to be relevant to the
same requests'' \cite{Rijsbergen:1979}. This hypothesis has been
typically used to improve the quality of the retrieval of documents
but there is very limited past work using the cluster hypothesis to
improve the quality of the evaluation.

Previous work on the expansion of an initial set of document
assessments include the use of Machine Learning. For example,
B\"{u}ttcher et al.~\cite{Buttcher2007} trained over a subset of qrels
in order to expand the set of qrels. They showed that evaluation
results with the expanded set of qrels had better quality than using
the source subset of qrels. Quality of the evaluation was measured by
ranking a set of IR systems according to the new expanded qrels, and
comparing it against the system ordering produced by the original
qrels.  In the clinical domain, Martinez et al.~\cite{martinez2008}
explored the use of re-ranking methods based on reduced judgements,
and found that the use of automatic classifiers would allow to
considerably reduce the time required for clinicians to identify a
large portion (95\%) of the relevant documents. Both of these articles
reported limitations of the classifiers when the initial number of
documents was small.
Furthermore, in the scenario that we contemplate, where we rely on the
list of references of a systematic review as the set of qrels, we do
not have information about negative qrels, and therefore a
classifier-based approach to expand the set of relevant documents
would have to deal with this issue.

More recent work~\cite{Sakai2010_PseudoQrel} has shown that by relying
on documents retrieved frequently by a diverse set of systems, it is
possible to build relevance assessments automatically, and achieve
high correlation with manually judged data. However this approach has
been tested by building on a set of competing runs from different
research groups, which is not always available; and this method does not
benefit from existing qrels. 

Prior work using document distance criteria for expanding the qrels
includes~\cite{Molla:2013}, who suggests that this approach may work
for a document collection within the medical domain. In this paper we
show that this approach improves the quality of evaluation \emph{both}
for medical and news reports, and we therefore add further evidence of
the plausibility of this method.

Our work complements that of related work on the study of the impact
of the number of topics and relevance judgements in IR
evaluation~\cite{Carterette:2007}.

\section{Data Sets}

We use the OHSUMED collection of medical research publications, and
the TREC-8 collection of news feeds.

The OHSUMED collection~\cite{Hersh1994} is a corpus containing
clinical queries and assessments. We focus on the set of 63 queries
that was used in the TREC-9 Filtering Track. The OHSUMED queries were
generated to address actual information needs for clinicians, and the
assessed documents were retrieved in two iterations, by relying on the
MEDLINE search interface\footnote{http://www.ncbi.nlm.nih.gov/pubmed}
and the SMART retrieval system respectively. The retrieved documents
were judged by a separate group of domain experts to the group
performing the search. As document collection we rely on the 1988-91
subset of MEDLINE that was released as test data for the TREC-9
challenge, which contains 293,856 documents. The judgement set has an
average of 50.87 judgements per query, all of them positive. Since the
original runs of the systems participating in the TREC-9 challenge are
not available, for evaluation we created 16 IR systems implemented
with the Terrier 3.5 open source
package~\cite{Macdonald2012_Terrier}. Table~\ref{tab:terrier} lists
the settings of the Terrier package used for our runs, which are the
same settings used by~\cite{Molla:2013}.
\begin{table}[t]
\centering
\scalebox{0.8}{
\begin{tabular}{|l | l | l | l |}
\hline
BB2 & BM25 & DFR\_BM25 & DLH \\
DPH & DFRee & Hiemstra\_LM & DLH13 \\
IFB2 & In\_expB2 & In\_expC2 & InL2 \\
LemurTF\_IDF & LGD & PL2 & TF\_IDF \\
\hline
\end{tabular}
}
\caption{List of 16 runs from the terrier package}
\label{tab:terrier}
\end{table}

Each document of the OHSUMED collection contains bibliographical data
(title, authors, etc) plus the abstract. For the experiments reported
in this paper we used only the contents of the abstract.

The TREC-8 collection~\cite{Voorhees:1999} comprises disks 4 and 5 of
the TREC collection, excluding the \emph{Congressional Record}
subcollection. We used the test set, which has 50 queries with an
average of 1,736 qrels per query. Of these, since we want to model a
scenario where only positive judgements are used, we use only the
positive qrels, which average 94.56 positive qrels per query. The
qrels were generated using the pooling method, taking the top 100
documents retrieved by the systems participating in the \emph{ad-hoc}
task of TREC-8. For evaluation we used the results of the original
systems that participated in the \emph{ad-hoc} track of TREC-8.

Each document of the TREC-8 collection contains various XML
markups. Given that each of the multiple sources had a different XML
tag set, for the experiments reported in this paper simply we ignored
all lines that had an XML markup. The remaining lines consisted mostly
of the main text, but there were still a few lines left that had
meta-data.

\section{Distance versus Relevance}\label{sec:distancerelevance}

We first examined the relation between similarity between qrel
candidates, and their relevance. We obtained the candidates by
pooling, as explained below for each dataset. For every query and for
every qrel candidate in the query, we computed the minimum distance
between the qrel candidate and a known positive qrel for the
query. The resulting (qrel candidate, query) pairs were sorted by
distance and binned into deciles such that the first decile is formed
by the top 10\% pairs, and so on. Then, within each decile we computed
the percentage of qrel candidates that were actually positive
qrels. Since the OHSUMED data only had positive qrels, for each query
we built the list of qrel candidates by pooling the top 100 documents
per run. There was an average of 202.80 qrel candidates per query
(12,371 qrel candidates in total\footnote{Note that the total number
  of qrels is slightly lower than 63*202.80=12,777 due to the
  existence of qrels shared among questions.}), and those that were
not in the list of known qrels were tagged as negative judgements. For
the TREC data, we used the qrels provided by the organisers of
TREC. These qrels had been obtained by pooling the top 100 documents
per run and contained positive and negative judgements, with an
average of 1,736.60 qrels per query (86,830 qrels in total). Due to
time and memory constraints we have used the first 100 qrels of each
query, giving a total of 5,000 qrel candidates.

\begin{figure}
  \centering
  \includegraphics[width=\columnwidth]{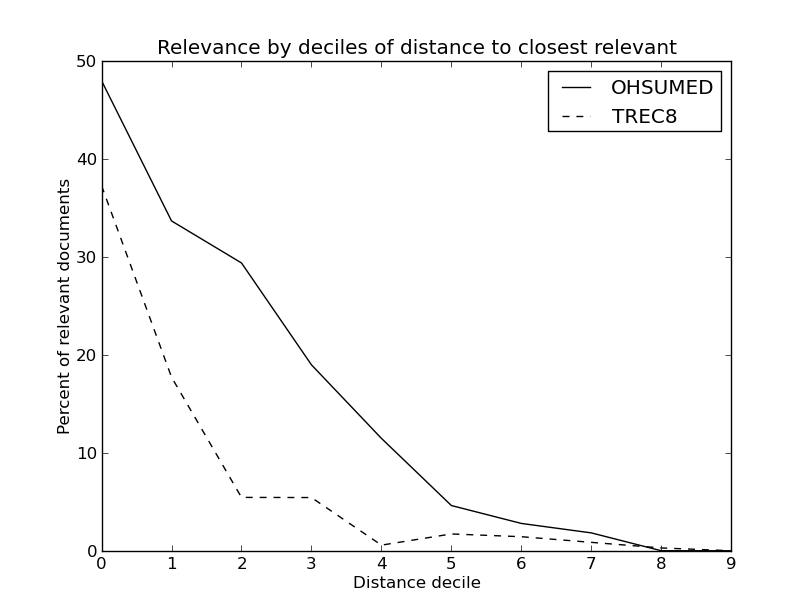}
  \caption{Distance versus relevance in the OHSUMED and TREC-8 test datasets.}
  \label{fig:distances}
\end{figure}

Figure~\ref{fig:distances} shows the result.
The figure shows a clear relation between distance and relevance in
both datasets. The relation is not as marked as reported by
\cite{Molla:2013} but, as we will show below, it is sufficient to give
an improvement in the evaluation when we expand the original
qrels. The reason why the results differ from those of prior work is
that the pool of documents in prior work was taken from the global
list of known qrels, instead of from the runs of the systems. Our
pooling method reflects a more realistic scenario and makes it
possible to compare the OHSUMED and the TREC datasets. We observe
that, in general, the percentage of relevant candidates drops much
quicker in the TREC data than in the OHSUMED data.

For the experiments we used as the distance metric
$d(x,y)=1-\cos(x,y)$ where $\cos(x,y)$ is the cosine similarity. The
vector representations were formed by obtaining the \emph{tf.idf}
values of all words after lowercasing and removing stop words, and
then taking the top 200 components after performing Principal
Component Analysis (PCA).\footnote{These experiments were carried out
  in Python and the scikit-learn library.} These are the same settings
as described by~\cite{Molla:2013}.

\subsection{Pseudo-qrels for Evaluation}

We expand the original qrels by introducing qrel candidates that are
close enough to a known positive qrel. The specific process to rank
the candidates is the same as described in
Section~\ref{sec:distancerelevance}. We then apply a percentile
threshold to select the pseudo-qrels. In other words, given the list
of pairs (qrel candidate, query) sorted by distance to the closest
positive qrel of the query, we select the top $K$\% qrel candidates.
We will call these added qrel candidates pseudo-qrels.

The process to find the pseudo-qrels uses a threshold that is global
to all queries. This means that some queries may receive more
pseudo-qrels than others, and a query may receive no pseudo-qrels. As
we reduce the threshold, we will find more cases where a query has no
additional pseudo-qrels. We thought that using a global threshold is
desirable, since if a query only has documents that are relatively far
from known qrels, we better not add them as pseudo-qrels.

To test the impact of the number of available qrels, in our
experiments we have varied the number of qrels per query, always
making sure that each query had at least one qrel. The selected qrels
were drawn randomly from the original set of qrels, using the same
random seed in all experiments.

\subsection{Correlation for ranking IR systems}

To determine the quality of the pseudo-qrels, and keeping in mind the
scenario envisaged at the introduction, we evaluate and rank the
set of runs using the qrels plus pseudo-qrels. The evaluation metric
was MAP. We then compare the ranking of systems against another
evaluation where we use the complete set of qrels. The system rankings
are compared using Kendall's tau.

We conducted several experiments by varying the percentages of qrels
extended with the computed pseudo-qrels. We also included a baseline
that does not include the additional pseudo-qrels. The baseline
simulates the default case when we only use the available qrels.

\begin{figure}[t]
  \centering
\includegraphics[width=\columnwidth]{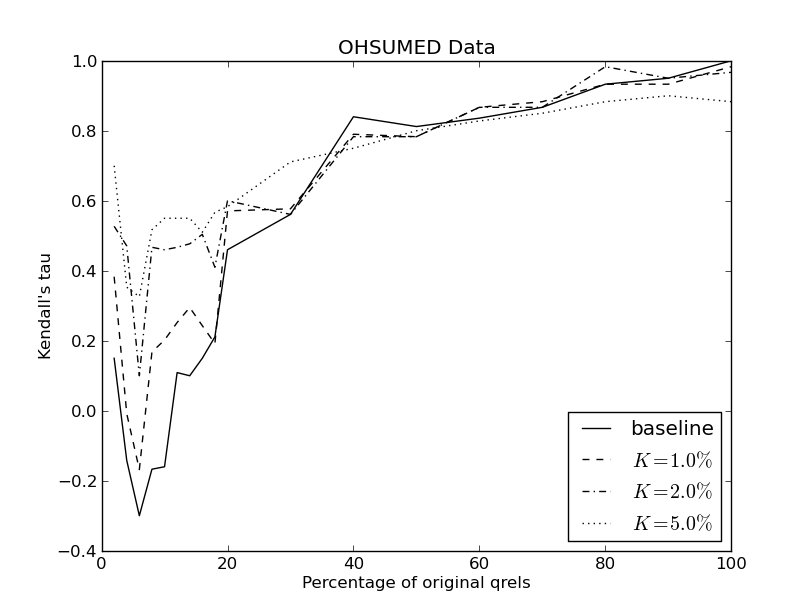}
\caption{Kendall's tau of system orderings on the OHSUMED data}
  \label{fig:results-ohsumed}
\end{figure}

\begin{figure}[t]
  \centering
\includegraphics[width=\columnwidth]{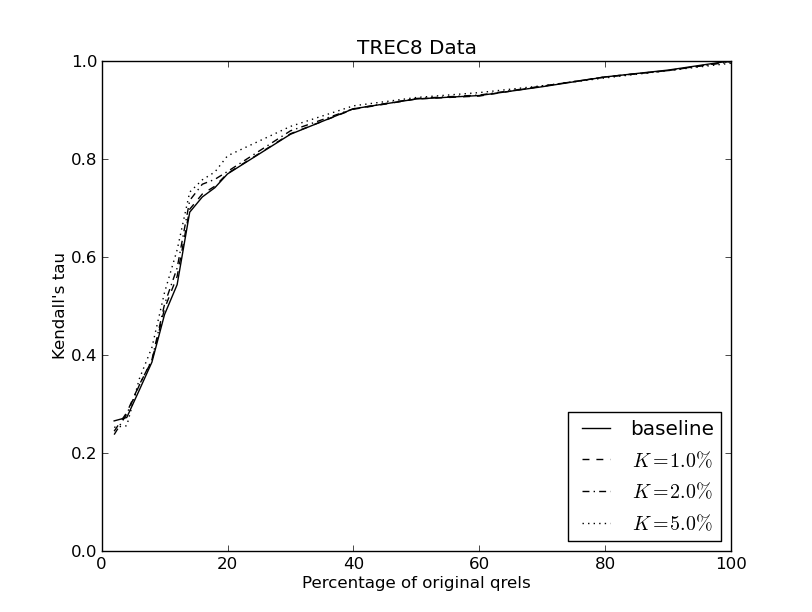}
\caption{Kendall's tau of system orderings on the TREC data}
  \label{fig:results-trec}
\end{figure}

Figure~\ref{fig:results-ohsumed} shows the results for the OHSUMED
dataset, and Figure~\ref{fig:results-trec} shows the results for the
TREC dataset. The figures present the results for varying values of
$K$ (the percentage of top documents selected as pseudo-qrels).  We
can observe, as expected, that larger percentages of qrels lead to
higher correlation.


\begin{figure}[t]
  \centering
\includegraphics[width=\columnwidth]{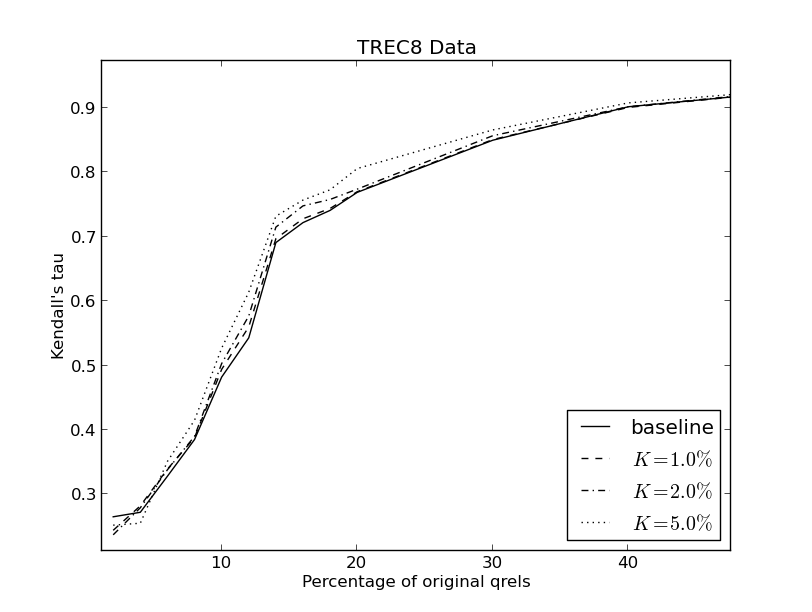}
\caption{Kendall's tau of system orderings focusing on the smaller percentages
  of the TREC data}
  \label{fig:results-trec-zoom}
\end{figure}

In both cases, we observe a gain of Kendall's tau for small
percentages $K$ of the original qrels. The gain is higher in the
OHSUMED than the TREC dataset. Figure~\ref{fig:results-trec-zoom}
zooms on the lower values of $K$ for the TREC data.
We appreciate a greater gain in some of the smaller values of
$K$. Critically, these values represent an original number of qrels
that is similar to those encountered in our envisaged
scenario.

\begin{figure}
  \centering
  \includegraphics[width=\columnwidth]{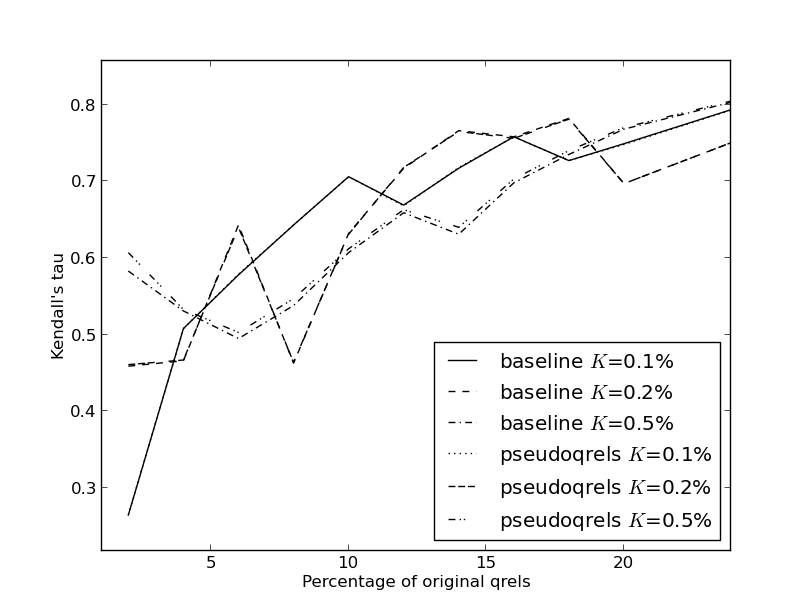}
  \caption{Impact of using different initial qrels. In all cases,
    adding pseudo-qrels improved the results or remained practically
    the same.}
  \label{fig:random}
\end{figure}

We observed that selecting a different subset of qrels influences the
resulting tau, especially for the smaller percentages of qrels. We
tried with several baselines by using different random seeds to select
the qrels, and compared them with the expanded versions with the
pseudo-qrels. The gain of adding pseudo-qrels varied depending on the
initial choice of qrels, but in general there was a
gain. Figure~\ref{fig:random} illustrates the impact of using
different initial qrels for the TREC dataset.

\section{Conclusions}

We have compared the use of document similarity scores in two
datasets, with the aim to compensate for the limited availability of
qrels. The advantage of our approach against classification-based
approaches such as those of prior work is that our method is
applicable even when there are only positive relevance judgements.

The results are particularly encouraging when the number of available
relevance judgements is very limited, and they suggest the use of
distance-metrics extensions of relevance judgements as a quick and
cheap evaluation step during the development stage of information
retrieval systems when there are few and only positive relevance
judgements. It can therefore be applied for the development of IR
systems that search for relevant clinical studies, even when the set
of known available relevant documents is just the list of references
of a sample clinical systematic review.

Further work includes a more comprehensive study of the thresholds
that lead to the best evaluation setting, and the use of variants of
distance metrics, other than straight cosine distance over a
bag-of-words vector space model. Also, given that the measure of
quality used in this study is based on the correlation of rankings
with an automated evaluation metric, it is desirable to extend this
study with real human judgements.

Finally, note that the present study expands the available qrels with
positive judgements only. A further interesting line of research will
include the automatic addition of negative judgements.

\section{Acknowledgments}

NICTA is funded by the Australian Government as represented by the
Department of Broadband, Communications and the Digital Economy and
the Australian Research Council through the ICT Centre of Excellence
program.

%
\bibliographystyle{abbrv}
\bibliography{sigir2014}  

\end{document}